\documentclass[runningheads]{llncs}
\usepackage[english]{babel}
\usepackage{subcaption}
\usepackage{xcolor}
\usepackage{url}
\usepackage{colortbl}
\usepackage{amssymb}
\usepackage{stmaryrd}
\usepackage{amsmath}
\usepackage{mathrsfs}
\usepackage{amssymb}
\usepackage[left]{lineno}
\usepackage{xfrac}
\usepackage{nicefrac}
\usepackage[textsize=scriptsize,backgroundcolor=yellow!40]{todonotes}
\usepackage[nonumberlist,acronym,sanitize=none]{glossaries}
\glsdisablehyper
\usepackage{comment}
\usepackage[pdftex, colorlinks=true, hyperfootnotes=true, hyperindex=true,
            plainpages=false, pagebackref=false, pdfpagelabels=true, pdfstartview=FitH,
            linkcolor=blue, citecolor=blue, urlcolor=blue,
            bookmarks, bookmarksopen, bookmarksdepth=3]{hyperref}
\usepackage[capitalise,nameinlink]{cleveref}
\crefname{algocf}{alg.}{algs.}
\Crefname{algocf}{Algorithm}{Algorithms}
\crefname{section}{Sect.}{Sects.} 
\crefname{Section}{Section}{Sections}
\usepackage{tkz-base}
\usetikzlibrary{decorations.pathmorphing,trees,snakes,arrows,shapes,automata,petri}
\usepackage{paralist}
\usepackage{multicol}
\usepackage{booktabs}
\usepackage{enumitem}
\usepackage{multirow}
\usepackage{rotating}
\usepackage{etaremune}
\usepackage[ruled,linesnumbered,algo2e]{algorithm2e}
\usepackage{marginnote}
\usepackage{mathtools}
\usepackage{mathabx}
\usepackage{adjustbox}
\usepackage{ifthen}
\usepackage[normalem]{ulem}
\usepackage{lineno}
\usepackage{soul}
\usepackage{floatrow}
\usepackage{listings}

\crefname{lstlisting}{Listing}{Listings}
\usepackage{lipsum}
\usepackage{makecell}
\usepackage{diagbox}
\usepackage[scientific-notation=false,group-separator={,}]{siunitx}
\usepackage{microtype}
\usepackage{footmisc}
\usepackage{wrapfig}
\usepackage{orcidlink}
\usepackage{textcase}
\usepackage[ddmmyyyy]{datetime}

\usepackage[T1]{fontenc}
\usepackage{lmodern}

\definecolor{specializedcliniccolor}{RGB}{0, 153, 0}
\definecolor{pharmacolor}{RGB}{0, 110, 175}
\definecolor{scriptcolor}{RGB}{237, 237, 237}
\newcommand{\Actor}[1]{\textrm{\MakeTextLowercase{#1}}}
\newcommand{\Compo}[1]{\texttt{#1}}
\newcommand{\Activ}[1]{\textrm{#1}}
\def\Org{\ensuremath{\mathscr{O}}}

\def\RCoefficent{\ensuremath{R^2}}
\newcommand{\RCoefficentF}[0]{\ensuremath{%
		 1-\frac{\sum_{i=1}^{k}(y_i- \hat{y_i})^2}{\sum_{i=1}^{k}(y_i- \overline{y_i})^2}}}
\def\Rlin{\ensuremath{R^2_\textrm{lin}}}
\def\Rlog{\ensuremath{R^2_\textrm{log}}}

\def\Slope{\ensuremath{\widehat{\mathcal{\beta}}}}
\newcommand{\SlopeF}[0]{\ensuremath{%
\frac{\sum_{i=1}^k\left(x_i-\bar{x}\right)\left(y_i-\bar{y}\right)}{\sum_{i=1}^k\left(x_i-\bar{x}\right)^2}}}

\def\OrgU {\ensuremath{\widehat{\mathcal{O}}}} 

\def\LPrv {\ensuremath{\mathcal{P}}}   
\def\LPrvF {\ensuremath{\mathfrak{p}}}   
\def\SecM {\ensuremath{\mathcal{M}}}   
\def\LPrvS {\ensuremath{\widehat{\mathcal{P}}}}  
\def\Case {\ensuremath{\textsl{C}}}  
\def\Evt {\ensuremath{e}}  
\def\EvtU {\ensuremath{\widehat{\mathrm{E}}}}  
\def\CasP {\ensuremath{\Case^{\LPrv}}}  
\def\CId {\ensuremath{\textrm{\textsl{cid}}}}  
\def\CIdF {\ensuremath{\mathfrak{id}}}  
\def\CIdU {\ensuremath{\widehat{\mathrm{CID}}}}  
\def\CasU {\ensuremath{\widehat{\mathrm{C}}}}  
\def\Merge {\ensuremath{\bigoplus}}

\def\SegSize {\ensuremath{\textsl{seg\_size}}}

\usepackage{newtxtext}

\usepackage[subtle]{savetrees}
\crefformat{enumi}{#2\textup{#1}#3}

\newenvironment{newj}[1][]{%
        \noindent
        \color{black}
 
}{%
	\normalcolor%
}
\newcommand\NewJ[1]{{\textcolor{black}{#1}}}
 
\begin{document}
%
\title{Trusted Execution Environment for \\ Decentralized Process Mining}
\titlerunning{TEE for Decentralized Process Mining}
%
\author{%
Valerio~Goretti\inst{1}\orcidlink{0000-0001-9714-4278} \and 
Davide~Basile\inst{1}\orcidlink{0000-0002-5804-4036} \and 
Luca~Barbaro\inst{1}\orcidlink{0000-0002-2975-5330} \and 
Claudio~Di~Ciccio\inst{1,2}\orcidlink{0000-0001-5570-0475}
}
\authorrunning{V.\ Goretti, D.\ Basile, L.\ Barbaro, C.\ Di Ciccio}
%
\institute{Sapienza University of Rome, Italy, \email{\href{mailto:davide.basile@uniroma1.it;luca.barbaro@uniroma1.it;valerio.goretti@uniroma1.it;claudio.diciccio@uniroma1.it}{name.surname@uniroma1.it}}
\and
Utrecht University, Netherlands, \email{\href{mailto:c.diciccio@uu.nl}{c.diciccio@uu.nl}}}
\maketitle

\begin{abstract}
 Inter-organizational business processes involve multiple independent organizations collaborating to achieve mutual interests. 
Process mining techniques have the potential to allow these organizations to enhance operational efficiency, improve performance, and deepen the understanding of their business based on the recorded process event data. 
However, inter-organizational process mining faces substantial challenges, including topical secrecy concerns: The involved organizations may not be willing to expose their own data to run mining algorithms jointly with their counterparts or third parties. 
In this paper, we introduce CONFINE, a novel approach that unlocks process mining on multiple actors' process event data while safeguarding the secrecy and integrity of the original records in an inter-organizational business setting.
To ensure that the phases of the presented interaction protocol are secure and that the processed information is hidden from involved and external actors alike, our approach resorts to a decentralized architecture comprised of trusted applications running in Trusted Execution Environments (TEEs). We show the feasibility of our solution by showcasing its application to a healthcare scenario and evaluating our implementation in terms of memory usage and scalability on real-world event logs.

\end{abstract}

\section{Introduction}
\label{sec:introduction}
In today's business landscape, organizations constantly seek ways to enhance operational efficiency, increase performance, and gain valuable insights to improve their processes. Process mining offers techniques to discover, monitor, and improve business processes by extracting knowledge from chronological records recorded by process-aware information systems, i.e., the \textit{event logs}~\cite{DBLP:books/sp/22/WeerdtW22}.
The vast majority of process mining contributions consider \textit{intra-organizational} settings, in which processes are executed inside individual organizations. However, organizations increasingly recognize the value of collaboration and synergy in achieving operational excellence. \textit{Inter-organizational} business processes involve several independent organizations cooperating to achieve a shared objective.
\NewJ{%
Process mining can bring the advantages of transparency, performance optimization, and benchmarking in this context~\cite{van2011intra}.
Since different process data owners feed separate mining nodes, this setting characterizes what we call \emph{decentralized process mining}.
%
}%
\begin{newj}
Companies, though, are reluctant to share private information required to execute process mining algorithms with external parties~\cite{liu2009challenges}, thus hindering its adoption. Letting sensitive operational data traverse organizational boundaries introduces concerns about data secrecy, security, and compliance with internal regulations~\cite{muller2021trust}. 
%
%
To address this issue, the majority of research endeavors have focused thus far on the alteration of input data or of intermediate analysis by-products, with the aim to impede the counterparts from reconstructing the original information sources~\cite{DBLP:journals/dke/FahrenkrogPetersenAW23,DBLP:conf/icpm/Fahrenkrog-Petersen19,elkoumy2020shareprom,elkoumy2020secure}. These preemptive solutions have the remarkable merit of neutralizing information leakage by malicious parties a priori. Nevertheless, they entail an ex-ante information loss, thus compromising downstream process mining capabilities~\cite{DBLP:journals/dke/FahrenkrogPetersenAW23,DBLP:conf/icpm/Fahrenkrog-Petersen19}, or require the execution of computationally heavy protocols undermining scalability~\cite{elkoumy2020shareprom,elkoumy2020secure}.

To overcome these limitations, we propose CONFINE, a novel approach and tool aimed at enhancing collaborative information system architectures with secrecy-preserving process mining capabilities. To secure information secrecy during the exchange and elaboration of data, our solution resorts to \emph{Trusted Execution Environments} (TEEs)~\cite{DBLP:conf/trustcom/SabtAB15}, namely hardware-secured contexts that guarantee code integrity and data confidentiality before, during, and after their utilization. Owing to these characteristics, CONFINE lets information be securely transferred beyond the organizations' borders. Therefore, computing nodes other than the information provisioners can aggregate and elaborate the original, unaltered process data in a secure, externally inaccessible vault. Also, CONFINE is capable of providing these guarantees while demanding scalable computational overhead. 

The decentralized architecture of CONFINE supports a four-staged protocol:
\end{newj}
\begin{inparaenum}[\itshape(i)\upshape]
	\item The initial exchange of preliminary metadata,
	\item the attestation of the mining entities,
	\item the secure transmission and secrecy-preserving merge of encrypted information segments amid multiple parties,
	\item the isolated and verifiable computation of process discovery algorithms on joined data.
\end{inparaenum}
We evaluate our proof-of-concept implementation against synthetic and real-world data with a convergence test followed by experiments to assess the scalability of our approach.
\NewJ{%
Since TEEs operate with dedicated memory pages shielded from access by external entities (operating system included), thus entailing a hardware constraint on computation space, we endow our experiments with an analysis of memory usage, too.
}

The remainder of this paper is as follows. \Cref{sec:background} provides an overview of related work. 
In \cref{sec:motivating}, we introduce a motivating use-case scenario in healthcare. We present the CONFINE approach in \cref{sec:design}. We describe the implementation of our approach in \cref{sec:realization}. 
In \cref{sec:evaluation}, we report on the efficacy and efficiency tests for our solution.
Finally, we conclude our work and outline future research directions in \cref{sec:conclusion}.

\section{Related Work}
\label{sec:background}
The scientific literature already includes noticeable contributions to process mining  
\NewJ{in a decentralized setting with a focus on data secrecy,}
despite the relative recency of this research branch across process mining and collaborative information systems. 
The work of M{\"u}ller et al.~\cite{muller2021process} revolves around data privacy and security within third-party systems that mine data generated from external providers on demand. To safeguard the integrity of data earmarked for mining purposes, their research introduces a conceptual architecture that entails the execution of process mining algorithms within a cloud service environment, fortified with Trusted Execution Environments. 
Drawing inspiration from this foundational contribution, our research work seeks to design a decentralized approach characterized by organizational autonomy in the execution of process mining algorithms, devoid of synchronization mechanisms taking place between the involved parties. A notable departure from the framework of M{\"u}ller et al.\ lies in the fact that here 
each participating organization retains the discretion to choose when and how mining operations are conducted. Moreover, we bypass the idea of fixed roles, engineering a peer-to-peer scenario in which organizations can simultaneously be data provisioners or miners. \NewJ{Fahrenkrog-Petersen et al.~\cite{DBLP:journals/dke/FahrenkrogPetersenAW23,DBLP:conf/icpm/Fahrenkrog-Petersen19} theorize the PRETSA algorithms family, namely a set of event log sanitization techniques that perform step-wise transformations of prefix-tree event log representation into a sanitized output ensuring \emph{k-anonimization} and \emph{t-closeness}. While these algorithms effectively minimize information loss, they introduce targeted approximations within the original event log, which may compromise the exactness of process mining results or inhibit mining tasks. 
In contrast, our research proposes an architecture wherein secure computational vaults collect event logs devoid of upstream alterations and protect them at runtime, thus generating results derived directly from the original information source.} Elkoumy et al.~\cite{elkoumy2020shareprom,elkoumy2020secure} present Shareprom. Like our work, their solution offers a means for independent entities to execute process mining algorithms in inter-organizational settings while safeguarding the proprietary input data from exposure to external parties operating within the same context.
Shareprom's functionality, though, is confined to the execution of operations involving event log abstractions~\cite{FederatedPM2021} represented as directed acyclic graphs, which the parties employ as intermediate pre-elaboration to be fed into secure multiparty computation (SMPC)~\cite{SMPC2015}. As the authors remark, 
relying on this specific graph representation imposes constraints that may turn out to be limiting in a number of process mining scenarios.
In contrast, our approach allows for the secure, ciphered transmission of event logs (or segments thereof) to process mining nodes. 
Moreover, SMPC-based solutions require computationally intensive operations and synchronous cooperation among multiple parties, which make these protocols challenging to manage as the number of participants scales up~\cite{SMPC2019}. In our research work, 
individual computing nodes run the calculations, 
thus not requiring synchronization with other machines once the input data is loaded. 

We are confronted with the imperative task of integrating event logs originating from different data sources and reconstructing 
consistent traces that describe collaborative process executions.
Consequently, we engage in an examination of 
methodologies delineated within the literature, each of which offers insights into the merging of event logs within inter-organizational settings.
The work of Claes et al.~\cite{claes2014merging} holds particular significance for our research efforts. Their seminal study introduces a two-step mechanism operating at the structured data level, contingent upon the configuration and subsequent application of merging rules. Each such rule indicates 
the relations between attributes of the traces and/or the activities that must hold across 
distinct traces 
to be combined. 
In accordance with their principles, our research incorporates a structured data-level merge based on case references and timestamps as merging attributes. The research by Hernandez et al.~\cite{hernandez2021merging} posits a methodology functioning at the raw data level. Their approach represents traces and activities as \textit{bag-of-words} vectors, subject to cosine similarity measurements to discern links and relationships between the traces earmarked for combination. An appealing aspect of this approach lies in its capacity to generalize the challenge of merging without necessitating a-priori knowledge of the underlying semantics inherent to the logs under consideration. However, it entails computational overhead in the treatment of data that can interfere with the overall effectiveness of our approach. 

\def\Talice{\Activ{PH, COPA, OD, \color{pharmacolor}DOR, PDL, SD, \color{black}RD, AD, TP, \color{specializedcliniccolor}PAFH, PIA, PT, VRT, TPB, \color{black}RPB, DPH, PCD, DP}}
\def\Tbob{\Activ{PH, COPA, OD, \color{pharmacolor}DOR, PDL, SD, \color{black}RD, AD, PRTA, PCD, DPH, DP}}
\def\TaliceUncolored{\Activ{PH, COPA, OD, DOR, PDL, SD, RD, AD, TP, PAFH, PIA, PT, VRT, TPB, RPB, DPH, PCD, DP}}
\def\TbobUncolored{\Activ{PH, COPA, OD, DOR, PDL, SD, RD, AD, PRTA, PCD, DPH, DP}}
\begin{figure}[t]
\centering
\includegraphics[width=\linewidth]{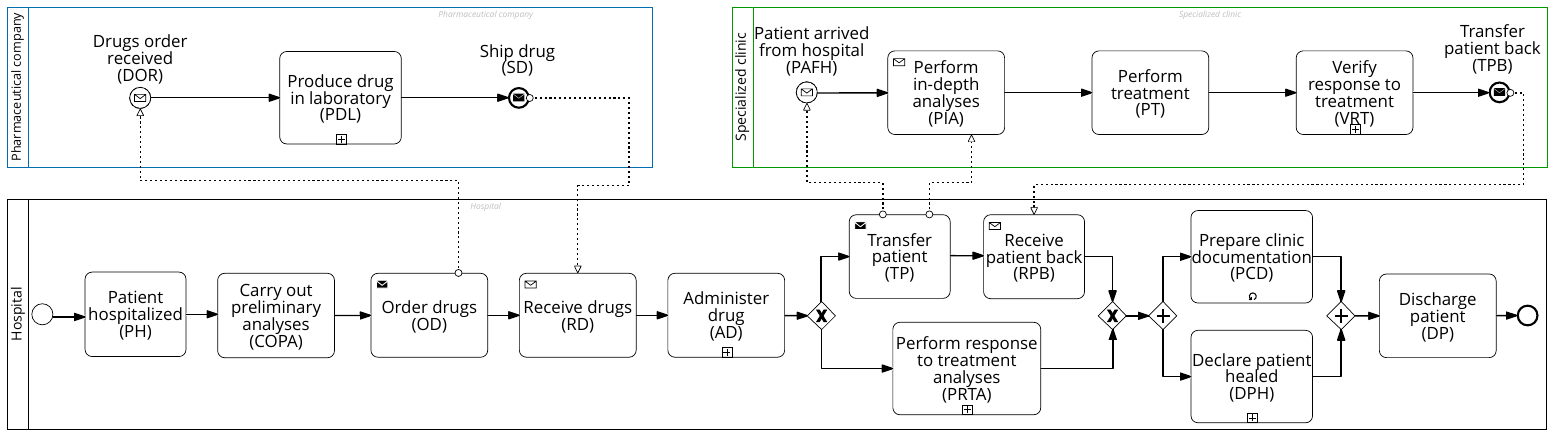}
\caption{A BPMN collaboration diagram of a simplified healthcare scenario}
\label{fig:BPMN_Healthcare}
\end{figure}
\begin{table}[t]
  \caption[Event log]{Events from cases 312 (Alice) and 711 (Bob) recorded by the \Actor{Hospital}, the \Actor{Specialized clinic}, and the \Actor{Pharmaceutical company}}
 \centering
    \begin{minipage}[t]{0.49\linewidth}
       \resizebox{0.98\textwidth}{!}{
        \begin{tabular}{|l|l|l||l|l|l|}
        \hline
        \multicolumn{6}{|c|}{Hospital}                                               \\ \hline
        {Case} & {Timestamp} & Activity & {Case} & {Timestamp} & Activity \\ \hline
        {312}         & {2022-07-14T10:36}           & PH    & {312}         & {2022-07-15T22:06}           & TP     \\ \hline
        {312}         & {2022-07-14T16:36}           & COPA  &   {711}         & {2022-07-16T00:55}           & PRTA          \\ \hline
        {711}         & {2022-07-14T17:21}           & PH    &   {711}         & {2022-07-16T01:55}           & PCD            \\ \hline
        {312}         & {2022-07-14T17:36}           & OD    &   {711}         & {2022-07-16T02:55}           & DPH           \\ \hline
        {711}         & {2022-07-14T23:21}           & COPA  &   {711}         & {2022-07-16T04:55}           & DP          \\ \hline
        {711}         & {2022-07-15T00:21}           & OD    &    {312}         & {2022-07-16T07:06}           & RPB        \\ \hline
        {711}         & {2022-07-15T18:55}           & RD    &    {312}         & {2022-07-16T09:06}           & DPH          \\ \hline
        {312}         & {2022-07-15T19:06}           & RD    &    {312}         & {2022-07-16T10:06}           & PCD         \\ \hline
        {711}         & {2022-07-15T20:55}           & AD    &    {312}        & {2022-07-16T11:06}           & DP            \\ \hline
        {312}         & {2022-07-15T21:06}           & AD     & \multicolumn{3}{c}{} \\ \cline{1-3}    
        \end{tabular}%
        }
    \end{minipage}%
    \hspace{0.001\textwidth}%
    \begin{minipage}[t]{0.49\linewidth}
        \resizebox{0.49\textwidth}{!}{
            \raisebox{2\baselineskip}{%
                \begin{tabular}{!{\color{pharmacolor}\vline}l!{\color{pharmacolor}\vline}l!{\color{pharmacolor}\vline}l!{\color{pharmacolor}\vline}}\arrayrulecolor{pharmacolor}
                \hline
                \multicolumn{3}{|c|}{Pharmaceutical company}                                               \\ \hline
                {Case} & {Timestamp} & Activity \\ \hline
                {312}         & {2022-07-15T09:06}           & DOR \\ \hline
                {711}         & {2022-07-15T09:30}           & DOR \\ \hline
                {312}         & {2022-07-15T11:06}           & PDL \\ \hline
                {711}         & {2022-07-15T11:30}           & PDL \\ \hline
                {312}         & {2022-07-15T13:06}           & SD \\ \hline
                {711}         & {2022-07-15T13:30}           & SD \\ \hline
                \end{tabular}%
            }
        }
        \hspace{0.001\textwidth}
        \resizebox{0.49\textwidth}{!}{
            \raisebox{2.5\baselineskip}{%
                \begin{tabular}{!{\color{specializedcliniccolor}\vline}l!{\color{specializedcliniccolor}\vline}l!{\color{specializedcliniccolor}\vline}l!{\color{specializedcliniccolor}\vline}}\arrayrulecolor{specializedcliniccolor}
                \hline
                \multicolumn{3}{!{\color{teal}\vline}c!{\color{teal}\vline}}{Specialized clinic}                                               \\ \hline
                {Case} & {Timestamp} & Activity \\ \hline
                {312}         & {2022-07-16T00:06}           & PAFH        \\ \hline
                {312}         & {2022-07-16T01:06}           & PIA         \\ \hline
                {312}         & {2022-07-16T03:06}           &  PT         \\ \hline
                {312}         & {2022-07-16T04:06}           & VRT         \\ \hline
                {312}         & {2022-07-16T05:06}           & TPB         \\ \hline
                \end{tabular}%
            }
        }
        \\[0.1\baselineskip]%
        \begin{minipage}[t]{\linewidth}
            \resizebox{\textwidth}{!}{%
                \begin{tabular}{r p{8cm}}
                $T_{312}=${\textlangle} & \Talice \,{\textrangle}\\[0.5\baselineskip]
                $T_{711}=${\textlangle} & \Tbob \,{\textrangle}\\
                \end{tabular}%
            }
        \end{minipage}
        \hfill
    \end{minipage}
     \label{tab:trace}
\end{table}

\section{Motivating Scenario}\label{sec:motivating}%
\NewJ{%
To provide a running example and motivating scenario for our investigation, we focus on a simplified hospitalization process for the treatment of rare diseases.
}%
The process model is depicted as a BPMN diagram in \cref{fig:BPMN_Healthcare} and involves the cooperation of three parties: a \Actor{Hospital}, a \Actor{Pharmaceutical company}, and a \Actor{Specialized clinic}.
For the sake of simplicity, we describe the process through two cases, recorded by the information systems as in \cref{tab:trace}. Alice's journey (\textbf{case 312}) begins when she enters the hospital for the preliminary examinations (patient hospitalized, \Activ{PH}). The \Actor{Hospital} then places an order for the drugs (\Activ{OD}) to the \Actor{Pharmaceutical company} for  treating Alice's specific condition. Afterwards, the \Actor{Pharmaceutical company} acknowledges that the drugs order is received (\Activ{DOR}), proceeds to produce the drugs in the laboratory (\Activ{PDL}), and ships the drugs (\Activ{SD}) back to the \Actor{Hospital}. Upon receiving the medications, the \Actor{Hospital} administers the drug (\Activ{AD}), and conducts an assessment to determine if Alice can be treated internally. If specialized care is required, the \Actor{Hospital} transfers the patient (\Activ{TP}) to the \Actor{Specialized clinic}. When the patient arrives from the hospital (\Activ{PAFH}), the \Actor{Specialized clinic} performs in-depth analyses (\Activ{PIA}) and proceeds with the treatment (\Activ{PT}). Once the \Actor{Specialized clinic} had completed the evaluations and verified the response to the treatment (\Activ{VRT}), it transfers the patient back (\Activ{TPB}). The \Actor{Hospital} receives the patient back \Activ(RPB) and prepares the clinical documentation (\Activ{PCD}). If Alice has successfully recovered, the \Actor{Hospital} declares the patient as healed (\Activ{DPH}). When Alice's treatment is complete, the \Actor{Hospital} discharges the patient (\Activ{DP}). 
Bob (\textbf{case 711}) enters the \Actor{Hospital} a few hours later. His hospitalization process is similar to Alice's. However, he does not need specialized care, and his case is only treated by the \Actor{Hospital}. Therefore, the \Actor{Hospital} performs the response to treatment analyses (\Activ{PRTA}) instead of transferring him to the \Actor{Specialized clinic}. 

Both the National Institute of Statistics of the country in which the three organizations reside and the University that hosts the hospital wish to uncover information on this inter-organizational process for reporting and auditing purposes via process analytics~\cite{Jans.Hosseinpour/IJAIS2019:ActiveLearningProcessMiningForAuditing}. The involved organizations share the urge for such an analysis and wish to be able to repeat the mining task also in-house. 
The \Actor{Hospital}, the \Actor{Specialized clinic}, and the \Actor{Pharmaceutical company} have a partial view of the overall unfolding of the inter-organizational process as they record the events stemming from the parts of their pertinence. 
In \cref{tab:trace}, we show cases 312 and 711 and the corresponding traces recorded by the \Actor{Hospital} (i.e., $T^H_{312}$ and $T^H_{711}$), the \Actor{Specialized clinic} (i.e., $T^S_{312}$ and $T^S_{711}$), and the \Actor{Pharmaceutical company} (i.e., $T^C_{312}$ and $T^C_{711}$).
Those traces are projections of the two combined ones for the whole inter-organizational process: $T_{312}=$\textlangle{}{\TaliceUncolored}\textrangle{} and $T_{711}=$\textlangle{}{\TbobUncolored}\textrangle{}. 
Results stemming from the analysis of the local cases would not provide a full picture. Data should be merged. However, to 
safeguard the confidentiality of the information, the involved parties cannot give other organizations open access to their traces. The diverging interests (being able to conduct process mining on data from multiple sources without giving away the local event logs in-clear) motivate our research. %
\begin{newj}
We remark that the problem we aim to solve spans across an array of domains beyond healthcare. It particularly applies to scenarios in which one or more parties are interested in process analytics outcomes based on data they bear but cannot be disclosed to the other process actors or to the miners.
In the supply chain realm, e.g., the extraction of aggregate knowledge about trends and management guidelines is called for, but the acquisition of competitive advantage out of knowledge leakage must be prevented~\cite{DBLP:journals/isf/TanWC16}. In personal informatics, company-wide work routine monitoring and analysis are desirable, though the details of individual participants should be sheltered from inquisitive inspections~\cite{DBLP:conf/rcis/SinikBR23}.
\end{newj}
\section{Design}\label{sec:design}
\NewJ{%
Our goal is to enable the secure aggregation and elaboration of original, unaltered event logs from decentralized sources in dedicated environments that potentially lie beyond the individual organizations' information perimeter.
With this objective in mind, we devise the \Compo{Secure Miner} component, which is capable of safeguarding data merge and processing by running certified code in an isolated execution vault. Thus, we decouple provisioning from treatment, and the two tasks can be carried out by distinct computing nodes.
Here, we introduce CONFINE's key components, with a special focus on the \Compo{Secure Miner}.} 
\begin{figure}[tb]
	\centering
	\includegraphics[width=0.75\linewidth]{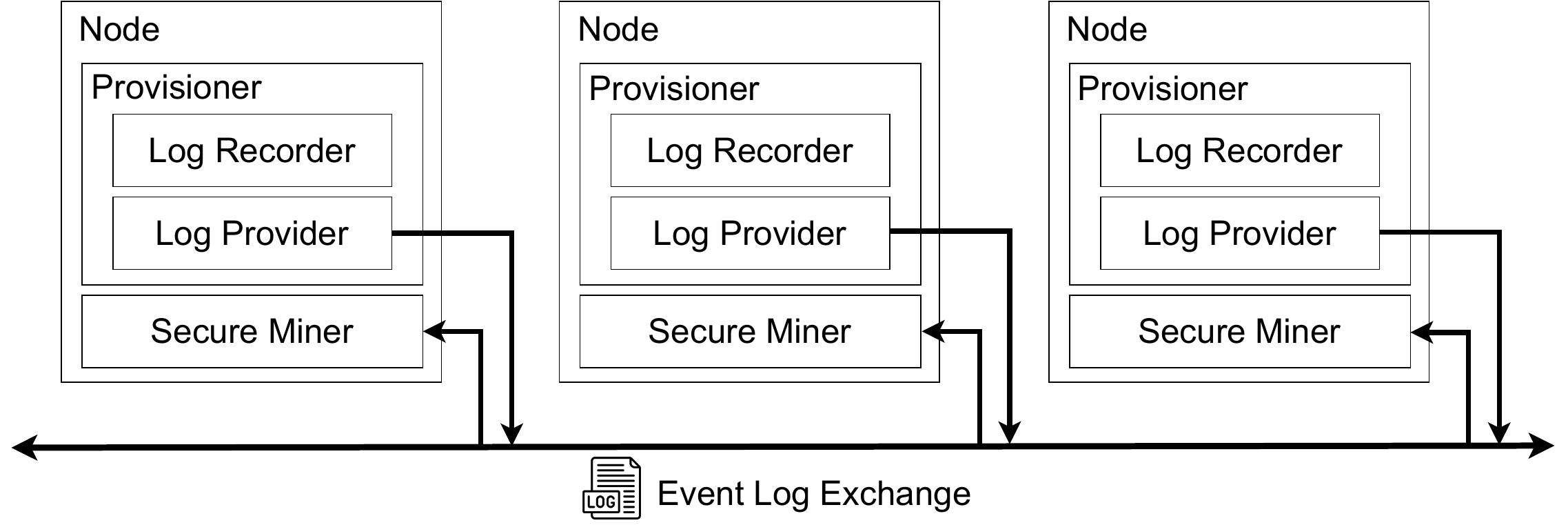}
	\caption{The CONFINE high-level architecture}
	\label{fig:architecture_diagram}
\end{figure} 

\noindent\textbf{The CONFINE architecture at large.}
Our architecture involves different information systems running on multiple machines. An organization can take at least one of the following roles: 
\begin{inparadesc}
\item[provisioning] if it delivers local event logs to be collaboratively mined;
\item[mining] if it applies process mining algorithms using event logs retrieved from provisioners.
\end{inparadesc}
\Cref{fig:architecture_diagram} depicts the high-level schematization of the CONFINE framework.
In our solution, each organization hosts one or more nodes encompassing diverse components (the names of which will henceforth be formatted with a \Compo{teletype} font). Depending on the played role, nodes come endowed with a \Compo{Provisioner} or a \Compo{Secure Miner}, or both. The \Compo{Provisioner} component, in turn, consists of the following two sub-components. The \begin{inparadesc}
\item[\Compo{Log Recorder}] registers the events taking place in the organizations' systems. The
\item[\Compo{Log Provider}] delivers on-demand data to miners.
\end{inparadesc}
The \Actor{Hospital} and all other parties in our example record Alice and Bob's cases using the \Compo{Log Recorder}. The \Compo{Log Recorder}, in turn, is queried by the \Compo{Log Provider} for event logs to be made available for mining. The latter controls access to local event logs by authenticating data requests by miners and rejecting those that come from unauthorized parties.
In our motivating scenario, the \Actor{Specialized clinic}, the \Actor{Pharmaceutical company}, and the \Actor{Hospital} leverage \Compo{Log Provider}s to authenticate the miner before sending their logs.  The \Compo{Secure Miner} component
shelters external event logs inside a protected environment to preserve data confidentiality and integrity.
Notice that \Compo{Log Provider}s accept requests issued solely by \Compo{Secure Miner}s. 
Next, we provide an in-depth focus on the latter.

\noindent\textbf{The Secure Miner.}
The primary objective of the \Compo{Secure Miner} is to allow miners to securely execute process mining algorithms using event logs retrieved from provisioners (the \Actor{Specialized clinic}, the \Actor{Pharmaceutical company}, and the \Actor{Hospital} in our example). \Compo{Secure Miner}s are isolated components that guarantee data inalterability and confidentiality. 

\begin{wrapfigure}[9]{r}{0.4\textwidth}
	\vspace{-1.5em}
	\centering
	\includegraphics[width=1\textwidth]{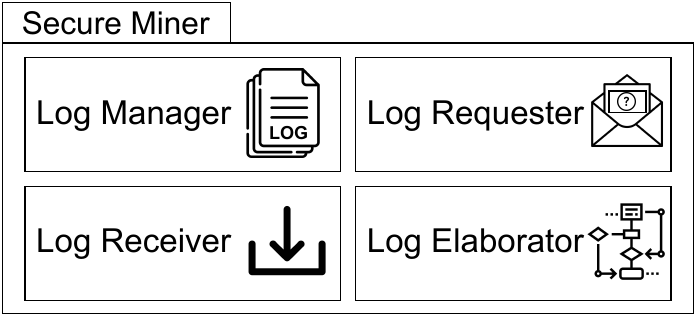}
	\caption[Secure Miner sub-components]{Sub-components of the \\Secure Miner}
	\label{fig:trusted_miner}
	\vspace{-6pt}
\end{wrapfigure}
\Cref{fig:trusted_miner} illustrates a schematization of the \Compo{Secure Miner}, which consists of four sub-components:
\begin{inparaenum}[\itshape(i)\upshape]
    \item the \Compo{Log Requester};
    \item the \Compo{Log Receiver};
    \item the \Compo{Log Manager}; 
    \item the \Compo{Log Elaborator}.
\end{inparaenum}
%
The \Compo{Log Requester} and the \Compo{Log Receiver} are the sub-components that we employ during the event log retrieval. \Compo{Log Requester}s send authenticable data requests to the \Compo{Log Provider}s. The \Compo{Log Receiver} collects event logs sent by \texttt{Log Providers} and entrusts them to the \Compo{Log Manager}, securing them from accesses that are external to the \Compo{Secure Miner}.
Miners of our motivating scenario, such as the \Actor{University} and the \Actor{National Institute of Statistics}, employ these three components to retrieve and store Alice and Bob's data. The \Compo{Log Manager} merges the event data locked in the \Compo{Secure Miner} to have a global view of the inter-organizational process comprehensive of activities executed by each involved party. The \Compo{Log Elaborator} executes process mining algorithms in a protected environment, inaccessible from the outside computation environment.
In our motivating scenario, the \Compo{Log Manager} combines the traces associated with the cases of Alice (i.e., $T^H_{312}$, $T^S_{312}$, and $T^C_{312}$) and Bob (i.e, $T^H_{711}$, $T^S_{711}$, and $T^C_{711}$), generates the chronologically sorted traces $T_{312}$ and $T_{711}$, and feeds them into the \Compo{Log Elaborator}'s mining algorithms (see the bottom-right quadrant of \cref{tab:trace}).

\section{Realization}
\label{sec:realization}
\NewJ{%
Thus far, we have outlined the main functionalities of each component at large. 
}%
Here we discuss the technical aspects concerning the realization of our solution. We 
first present the technologies through which we enable the design principles in \cref{sec:design}. Then, we discuss the CONFINE interaction protocol. Finally, we show the implementation details.

%
\begin{figure}[t]
	\centering
	\includegraphics[width=1\linewidth]{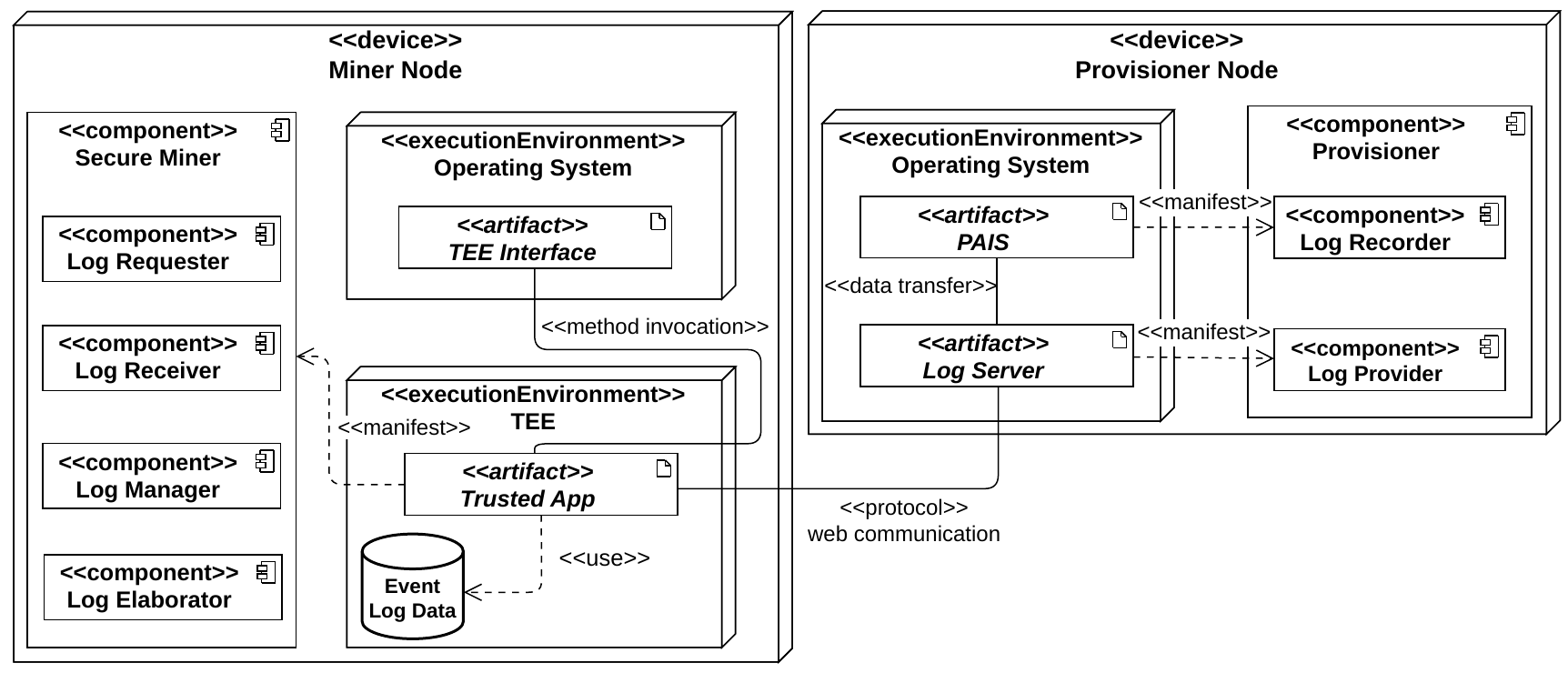}
	\caption{UML deployment diagram of the CONFINE architecture}
	\label{fig:deployment_diagram}
\end{figure}
\subsection{Deployment}
\label{sec:deployment}
%
\Cref{fig:deployment_diagram} depicts a UML deployment diagram~\cite{koch2002expressive} to illustrate the employed technologies and computation environments. 
We recall that the \Compo{Miner} and \Compo{Provisioner} nodes are drawn as separated, although organizations can host both.
In our motivating scenario, e.g., the \Actor{Hospital} can be equipped with machines aimed for both mining and provisioning. 

\Compo{Provisioner Node}s host the \Compo{Provisioner}'s components, i.e., the \Compo{Log Recorder} and the \Compo{Log Provider}. 
The Process-Aware Information System (\Compo{PAIS}) manifests the \Compo{Log Recorder}~\cite{Dumas.etal/2018:FundamentalsofBPM}. 
The \Compo{PAIS} grants access to the \Compo{Log Server}, enabling it to retrieve event log data. The \Compo{Log Server}, on the other hand, embodies the functionalities of the \Compo{Log Provider}, implementing services that handle remote data requests and provide event log data to the miners. 
%
The \Compo{Miner Node} is characterized by two distinct \textit{execution environments}: the \Compo{Operating System} (\Compo{OS}) and the Trusted Execution Environment (\Compo{TEE})~\cite{DBLP:conf/trustcom/SabtAB15}. \Compo{TEE}s establish isolated contexts separate from the \Compo{OS}, safeguarding code and data through hardware-based encryption mechanisms. This technology relies on 
dedicated sections of a CPU that handle encrypted data within a reserved section of the main memory~\cite{costan2016intel}.
\begin{newj}
By enforcing memory access restrictions, TEEs aim to prevent one application from reading or altering the memory space of another, thus enhancing system security.
These dedicated areas in memory are limited, though.
Once the limits are exceeded, TEEs have to scout around in outer memory areas, thus conceding the opportunity to malicious readers to understand the saved data based on the reads and writes.
To avoid this risk, TEE implementations often raise errors that halt the program execution when memory demand goes beyond the available space. 
Therefore, the design of secure systems that resort to TEEs must take into account that memory consumption must be kept under control.
\end{newj}

We leverage the security guarantees provided by TEEs~\cite{DBLP:journals/ieeesp/JauernigSS20} to protect a \Compo{Trusted App} responsible for fulfilling the functions of the \Compo{Secure Miner} and its associated sub-components. 
Our \Compo{TEE} component ensures the integrity of the \Compo{Trusted App} code, protecting it against potential malicious manipulations and unauthorized access by programs running within the \Compo{OS}. Additionally, we utilize the isolated environment of the \Compo{TEE} to securely store event log data (e.g., Alice and Bob's cases). 
The \Compo{TEE} retains a private key in its inaccessible memory section, paired with a public key in a Rivest-Shamir-Adleman
(RSA)~\cite{DBLP:journals/cacm/RivestSA83} scheme for attestation (only the owner of the private key can sign messages in a way that is verifiable via the public key) and secure message encryption (only the owner of the private key can decode messages that are encrypted with the corresponding public key).
In our solution, access to data located in the \Compo{TEE} is restricted to the sole \Compo{Trusted App}. Users interact with the \Compo{Trusted App} through the \Compo{TEE Interface}, which serves as the exclusive communication channel. The \Compo{Trusted App} offers secure methods, invoked by the \Compo{Trusted App Interface}, for safely receiving information from the \Compo{OS} and outsourcing the results of computations. 

\subsection{The CONFINE protocol}\label{sec:confine-protocol}
\begin{figure}[t]
	\subfloat[][Initialization]{\includegraphics[width=0.31\linewidth]{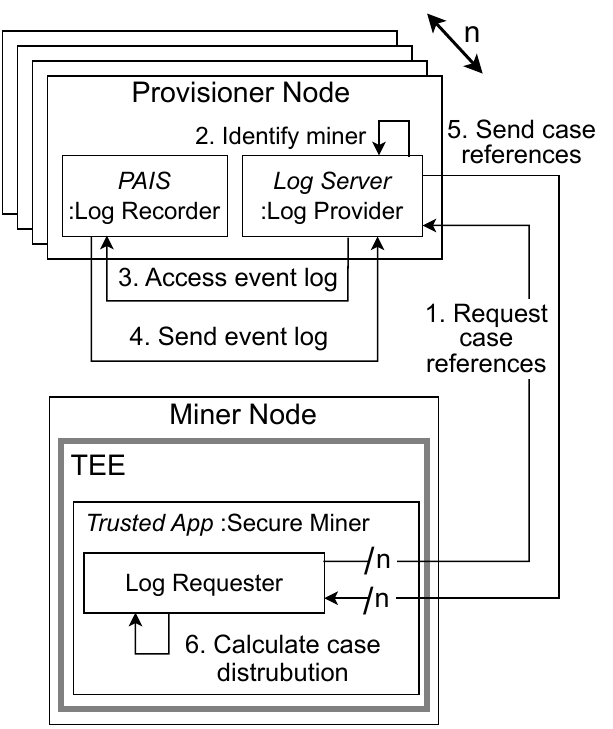}\label{fig:init}}\hfill
	\subfloat[][Remote attestation]{\includegraphics[width=0.32\linewidth]{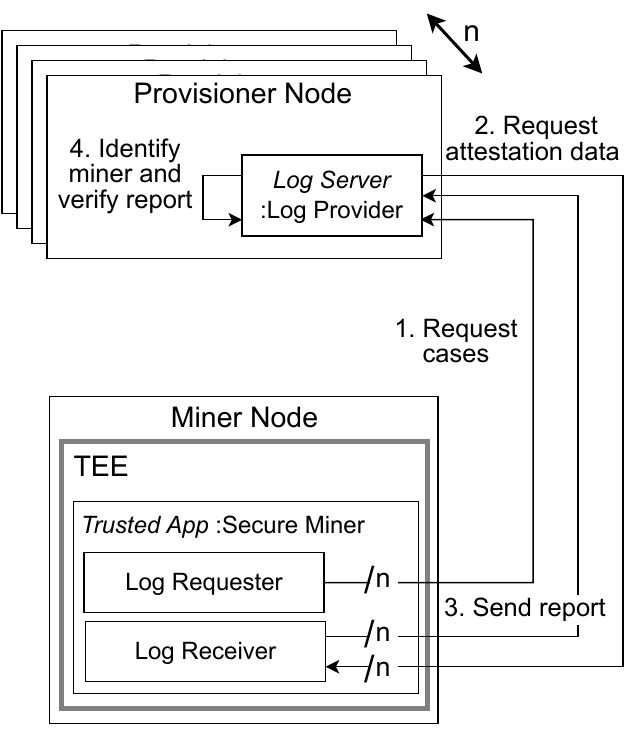}\label{fig:attestation}}\hfill
	\subfloat[][Data transmission]{\includegraphics[width=0.32\linewidth]{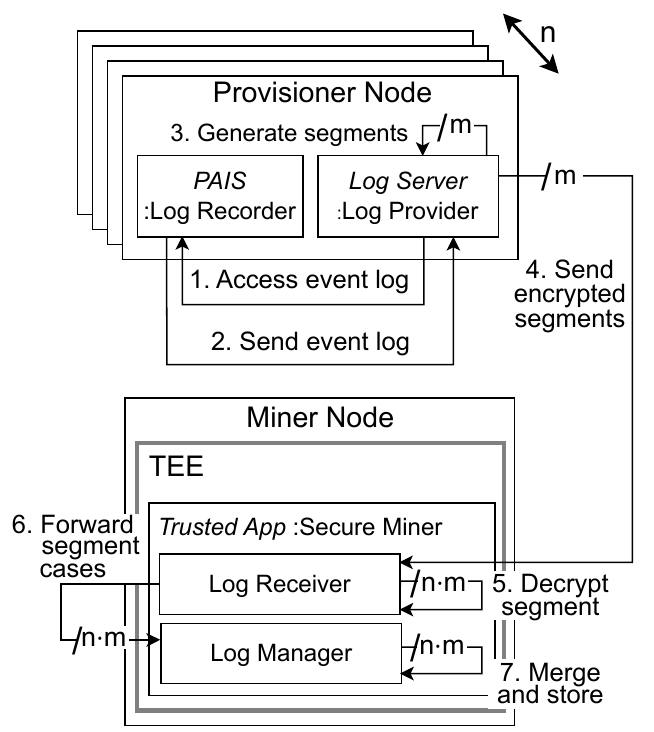}\label{fig:transmission}}\hfill
	\caption{Unfolding example for the initialization, remote attestation and data transmission phases of the CONFINE protocol}
	\label{fig:workflow}
\end{figure}

\begin{wrapfigure}[12]{r}{0.33\textwidth}
	\vspace{-1em}
	\includegraphics[width=\textwidth]{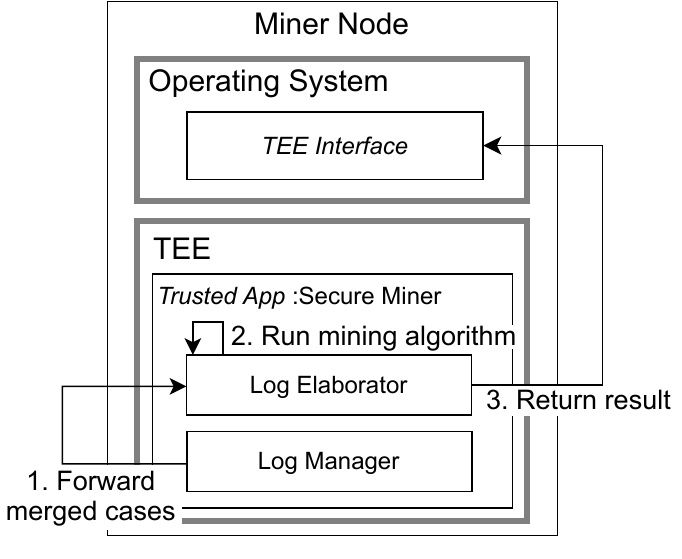}
	\caption[A gull]{Computation phase of the CONFINE protocol}
	\vspace{-2pt}
	\label{fig:computation}
\end{wrapfigure}

We orchestrate the interaction of the components in CONFINE via a protocol, which consists of four subsequent stages:
\begin{inparaenum}[\itshape(i)\upshape]
	\item \textit{initialization}, \item \textit{remote attestation}, \item \textit{data transmission}, and \item \textit{computation}.
\end{inparaenum}
These stages are depicted in \cref{fig:init,fig:attestation,fig:transmission,fig:computation}, respectively. They are mainly enacted by a \Compo{Miner Node}
\NewJ{%
(multiple instances of which can be deployed in a decentralized fashion)
}%
and $n$ \Compo{Provisioner Node}s. We assume their communication channel is reliable~\cite{Cachin.etal/2011:ReliableSecureDistributedProgramming} and secure~\cite{KamilLowe/FAST2010:AbstractionsSecureChannels}. In the following, we describe each of the above phases in detail.

\noindent\textbf{Initialization.} The objective of the initialization stage is to inform the miner about the distribution of cases related to a business process among the \Compo{Provisioner Nodes}. At the onset of this stage, the \Compo{Log Requester} within the \Compo{Trusted App} issues $n$ requests, one per \Compo{Log Server} component, 
to retrieve the list of case references they record (step 1 in \cref{fig:init}). Following sender authentication (2), each \Compo{Log Server} retrieves the local event log from the \Compo{PAIS} (3, 4) and subsequently responds to the \Compo{Log Requester} by providing a list of its associated case references (5). After collecting these $n$ responses, the \Compo{Log Requester} delineates the distribution of cases. In the context of our motivating scenario, by the conclusion of the initialization, the miner gains knowledge that the case associated with Bob, synthesized in the traces $T^H_{711}$ and $T^C_{711}$, is exclusively retained by the \Actor{Hospital} and the \Actor{Specialized Clinic}. In contrast, the traces of Alice's case, denoted as $T^H_{312}$, $T^C_{312}$, and $T^S_{312}$, are scattered across all three organizations.

\noindent\textbf{Remote attestation.} The remote attestation serves the purpose of establishing trust between miners and provisioners in the context of fulfilling data requests. This phase adheres to the overarching principles outlined in the RATS RFC standard~\cite{rfc9334} serving as the foundation for several TEE attestation schemes (e.g., Intel EPID%
\footnote{\url{sgx101.gitbook.io/sgx101/sgx-bootstrap/attestation}. Accessed: \today.}
and AMD SEV-SNP%
\footnote{\url{amd.com/en/processors/amd-secure-encrypted-virtualization}. Accessed: \today.}%
). Remote attestation has a dual objective:
\begin{inparaenum}[\itshape(i)\upshape]
    \item to furnish provisioners with compelling evidence that the data request for an event log originates from a \Compo{Trusted App} running within a TEE;
    \item to confirm the specific nature of the \Compo{Trusted App} as an authentic \Compo{Secure Miner} software entity.
\end{inparaenum}
This phase is triggered when the \Compo{Log Requester} sends a new case request to the \Compo{Log Server}, specifying:
\begin{inparaenum}[\itshape(i)\upshape]
    \item the segment size (henceforth, \SegSize), and
    \item the set of the requested case references.
\end{inparaenum}
Both parameters will be used in the subsequent \textit{data transmission} phase. Each of the $n$ \Compo{Log Server}s commences the verification process by requesting the necessary information from the \Compo{Log Receiver} to conduct the attestation (2). Subsequently, the \Compo{Log Receiver} generates the attestation report containing the so-called \emph{measurement} of the \Compo{Trusted App}, which is defined as the hash value of the combination of its source code and data. Once this report is signed using the attestation private key associated with the \Compo{TEE}'s hardware of the \Compo{Miner Node}, it is transmitted by the \Compo{Log Receiver} to the \Compo{Log Server}s alongside the attestation public key of the \Compo{Miner Node} (3). The \Compo{Log Server}s authenticate the miner using the public key and decrypt the report (4). In this last step, the \Compo{Log Server}s undertake a comparison procedure in which they juxtapose the measurement found within the decrypted report against a predefined reference value associated with the source code of the \Compo{Secure Miner}. If the decrypted measurement matches the predefined value, the \Compo{Miner Node} gains trust from the provisioner.

\noindent\textbf{Data transmission.} Once the trusted nature of the \Compo{Trusted App} is verified, the \Compo{Log Server}s proceed with the transmission of their cases. To accomplish this, each \Compo{Log Server} retrieves the event log from the \Compo{PAIS} (steps 1 and 2, in \cref{fig:transmission}), and filters it according to the case reference set specified by the miner. 
\NewJ{%
Given the constrained workload capacity of the \Compo{TEE}, 
\Compo{Log Server}s 
could be requested to partition the filtered event log into $m$ distinct segments. 
Log segments contain a variable count of entire cases (3).
}%
The cumulative size of these segments is governed by the threshold parameter specified by the miner in the initial request (step 1 of the remote attestation phase, \cref{fig:attestation}). As an illustrative example from our motivating scenario, the \Compo{Log Server} of the \Actor{Hospital} may structure the segmentation such that $T^H_{312}$ and $T^H_{711}$ are in the same segment, whereas the \Actor{Specialized clinic} might have $T^S_{312}$ and $T^S_{711}$ in separate segments. Subsequently, the $n$ \Compo{Log Server}s transmit their $m$ encrypted segments to the \Compo{Log Receiver} of the \Compo{Trusted App} (4). The \Compo{Log Receiver}, in turn, collects the $n \times m$ responses in a queue, processing them one at a time. After decrypting a processed segment (5), the \Compo{Log Receiver} forwards the cases contained therein to the \Compo{Log Manager} (6). 
Data belonging to the same process instance 
are merged by the \Compo{Log Manager} 
to build a single 
trace (e.g., $T_{312}$) 
comprehensive of all the events in the partial traces ($T^H_{312}$, $T^S_{312}$ and $T^C_{312}$). 
To do so, the \Compo{Log Manager} applies a specific \textit{merging schema} (i.e., a rule specifying the attributes that 
identify a case) as stated in~\cite{claes2014merging}. In our illustrative scenario, the merging schema to combine the cases of Alice is contingent upon the linkage established through their case identifier (312). We underline that our 
solution facilitates the incorporation of diverse merging schemas encompassing distinct trace attributes. The outcomes arising from merging the cases within the processed segments are securely stored by the \Compo{Log Manager} in the \Compo{TEE}.

\noindent\textbf{Computation.} The \texttt{Trusted App} requires all the provisioners to have delivered data referring to the same process instances. For example, when the \Actor{Hospital} and the other organizations have all delivered their information concerning case 312 to the \Compo{Trusted App}, the process instance associated with Alice becomes eligible for the computation phase, illustrated in \cref{fig:computation}. 
The \Compo{Log Manager} forwards the \NewJ{cases} earmarked for computation to the \Compo{Log Elaborator} (step 1).
\NewJ{%
These cases may constitute either the entire merged event log or a subset thereof. The former setting entails a single computation routine, thus saving execution time but requiring a larger memory buffer in the \Compo{TEE}, whereas the latter necessitates multiple consecutive elaborations with a lower demand for space.%
}   
Subsequently, the \Compo{Log Elaborator} proceeds to input the merged cases into the process mining algorithm (2). 
\NewJ{%
Notice that the above choice on the buffering of cases affects the selection of the mining algorithm to employ.
If we elaborate subsequent batches, each containing a part of all merged cases, the mining algorithm must support incremental processing, enriching the output as new batches come along. An example of this class of algorithms is the HeuristicsMiner~\cite{weijters2006process}.
Otherwise, incrementality is not required.
}
Ultimately, the outcome of the computation is relayed by the \Compo{Log Elaborator} from the \Compo{TEE} to the \Compo{TEE Interface} running atop the \Compo{Operating System} of the \Compo{Miner Node} (3). In our motivating scenario, the \Actor{University} and the \Actor{National Institute of Statistics}, serving as miners, disseminate the outcomes of computations, generating analyses that benefit the provisioners, although the original data are never revealed in clear. Furthermore, our protocol enables the potential for provisioners to have their own \Compo{Secure Miner}, allowing them autonomous control over the computed results. Notice that the CONFINE protocol does not impose restrictions on the post-computational handling of results.

%
\subsection{Implementation}
\label{sec:implementation:details}
We implemented the \texttt{Secure Miner} component as an Intel SGX%
\footnote{\url{sgx101.gitbook.io/sgx101/}. Accessed: \today.}
trusted application, encoded in Go through the EGo framework.%
\footnote{\url{docs.edgeless.systems/ego}. Accessed: \today.}
We resort to a TLS communication channel~\cite{Thomas/2000:SSL-TLS} between miners and provisioners over the HTTP web protocol to secure the information exchange. To demonstrate the effectiveness of our framework, 
we re-implemented and integrated the HeuristicsMiner discovery algorithm~\cite{weijters2006process} within the \Compo{Trusted Application}. 
Our implementation of CONFINE, including the HeuristicsMiner implementation in Go, is openly accessible at the following URL: \href{https://github.com/Process-in-Chains/CONFINE/}{\nolinkurl{github.com/Process-in-Chains/CONFINE/}}.

\label{sec:discussion:subsec:convergence}
\section{Evaluation}
\label{sec:evaluation}

In this section, we evaluate our approach through our implementation. 
We begin with a convergence analysis 
to demonstrate the correctness of the data exchange process. 
\NewJ{%
As discussed in~\cref{sec:deployment}, the availability of space in the dedicated TEE areas is subject to hardware limitations. Therefore, we focus on memory consumption, as exceeding those limits could lower the level of security guaranteed by TEEs.
}%
Thus, we 
gauge the memory usage with synthetic and real-life event logs, to observe the trend during the enactment of our protocol and assess scalability. We discuss our experimental results in the following.
All the testbeds and results are available in our public code repository (linked above).
\begin{figure}[bt]
	\includegraphics[width=1\linewidth]{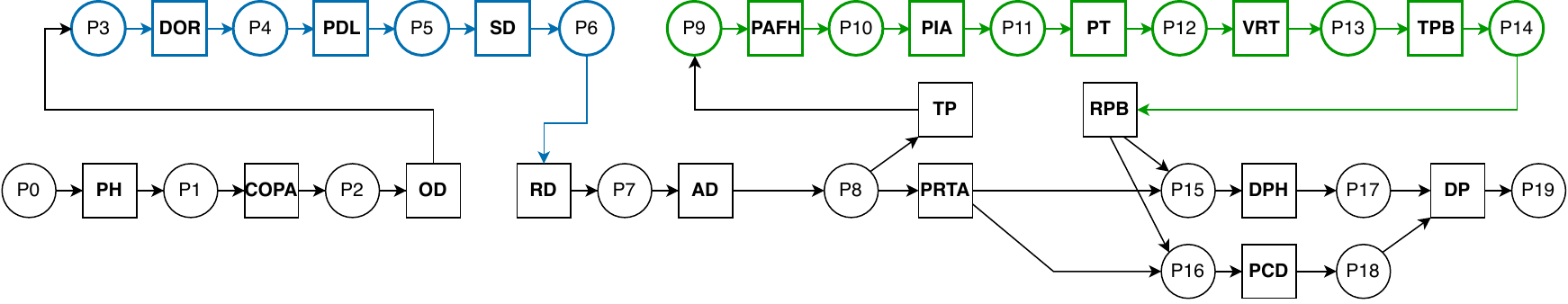}\label{fig:wfnet:d}
	\caption[HeuristicsMiner output]{HeuristicsMiner output with CONFINE}
	\label{fig:wfnet}
\end{figure}

\noindent\textbf{Output convergence.}
\label{sec:evaluation:subsec:convergence}
To experimentally validate the correctness of our approach in the transmission and computation phases (see \cref{sec:realization}), we run a \emph{convergence} test. 
To this end, we created a synthetic event log consisting of \num{1000} cases of \num{14} events on average (see \cref{tab:testedlogs}) by simulating the inter-organizational process of our motivating scenario 
(see \cref{fig:BPMN_Healthcare})%
\footnote{We generated the event log through BIMP (\url{https://bimp.cs.ut.ee/}). We filtered the generated log by keeping the sole events that report on the completion of activities, and removing the start and end events of the \Actor{Pharmaceutical company} and \Actor{Specialized clinic}'s sub-processes.}
and we partitioned it in three sub-logs (one per involved organization), an excerpt of which is listed in~\cref{tab:trace}.
We run the stand-alone HeuristicsMiner on the whole log, and processed the sub-logs through our CONFINE toolchain.
As expected, the results converge and are depicted in \cref{fig:wfnet} in the form of a workflow net~\cite{Aalst/ICATPN97:VerificationofWfNs}. For clarity, we have colored activities recorded by the organizations following the scheme of~\cref{tab:testedlogs} (black for the \Actor{Hospital}, blue for the \Actor{Pharmaceutical company}, and green for the \Actor{Specialized clinic}).
\begin{table}[t]
	\caption{Event logs used for our experiments}
	\label{tab:testedlogs}
	\centering
       \resizebox{\textwidth}{!}{%
            \begin{tabular}{l l S[table-format = 2.0] S[table-format = 4.0] S[table-format = 3.0] S[table-format = 1.0] S[table-format = 2.0] l} \toprule
            	\textbf{Name}               & \textbf{Type} & \textbf{Activities} & \textbf{Cases} & \textrm{\textbf{Max} events} & \textrm{\textbf{Min} events} & \textrm{\textbf{Avg}.\ events} & \textbf{Organization $\mapsto$ Activities}                        \\ \midrule
            	Motivating scenario  & Synthetic     & 19                  & 1000           &18 &9 &14 & ${\Org}^P \mapsto 3$, ${\Org}^C \mapsto 5$, ${\Org}^H \mapsto 14$ \\
            	Sepsis~\cite{seps}                       & Real          & 16                & 1050          & 185 & 3 & 15  & ${\Org}^1 \mapsto 1$, ${\Org}^2 \mapsto 1$, ${\Org}^3 \mapsto 14$ \\                                                                 
            	BPIC~2013~\cite{bpic2013}                   & Real          & 7                  & 1487           &123 &1 &9 & ${\Org}^1 \mapsto 6$, ${\Org}^2 \mapsto 7$, ${\Org}^3 \mapsto 6$  \\ \bottomrule& & 
            \end{tabular}
        }
\end{table}
\begin{figure}[t]
\centering
\begin{subfigure}{0.49\textwidth}
  \centering
  \includegraphics[width=\textwidth]{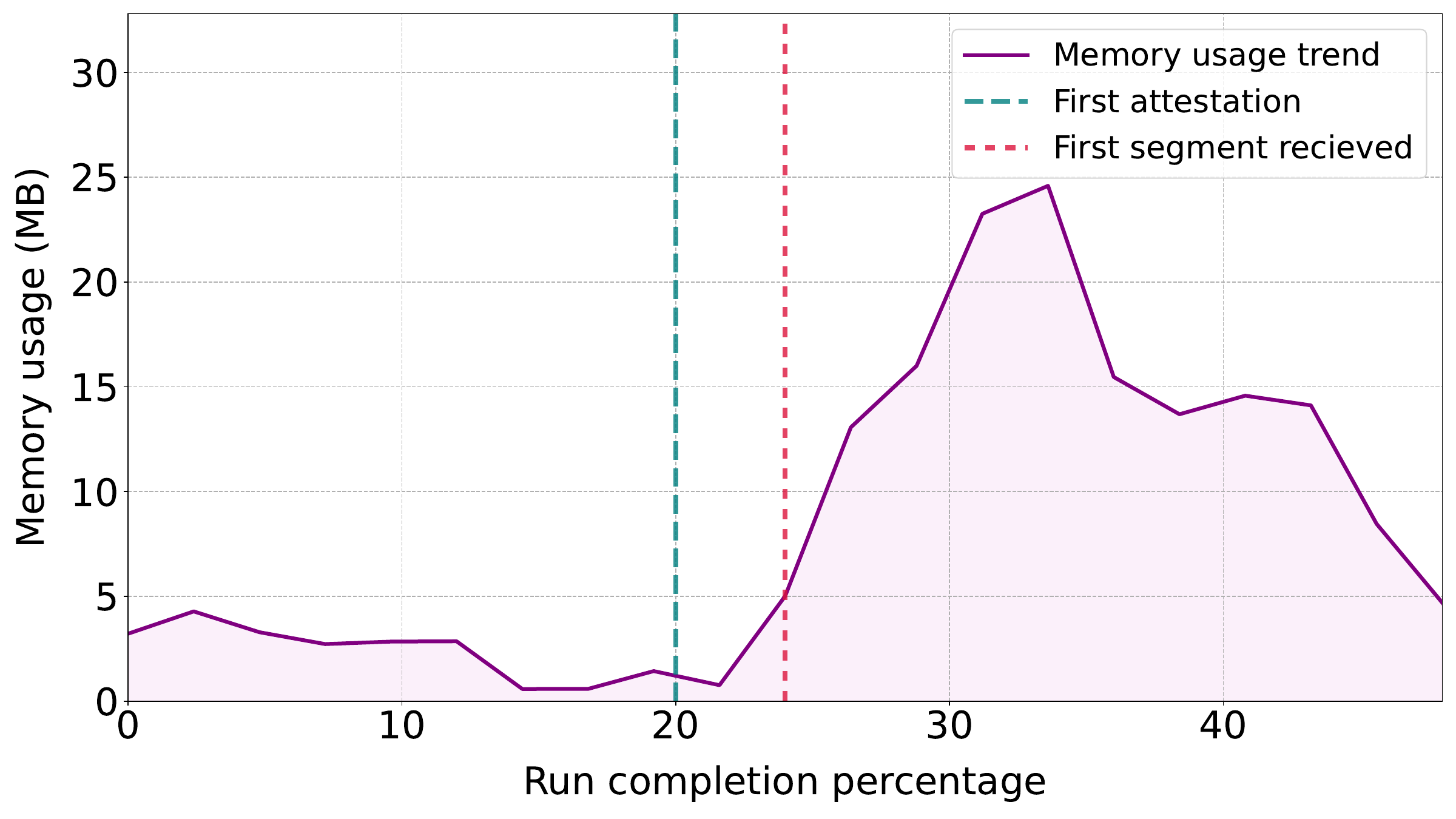}
  \caption{Memory usage without the computation phase}
  \label{snr_a}
\end{subfigure}\hfill
\begin{subfigure}{0.49\textwidth}
  \centering
  \includegraphics[width=\textwidth]{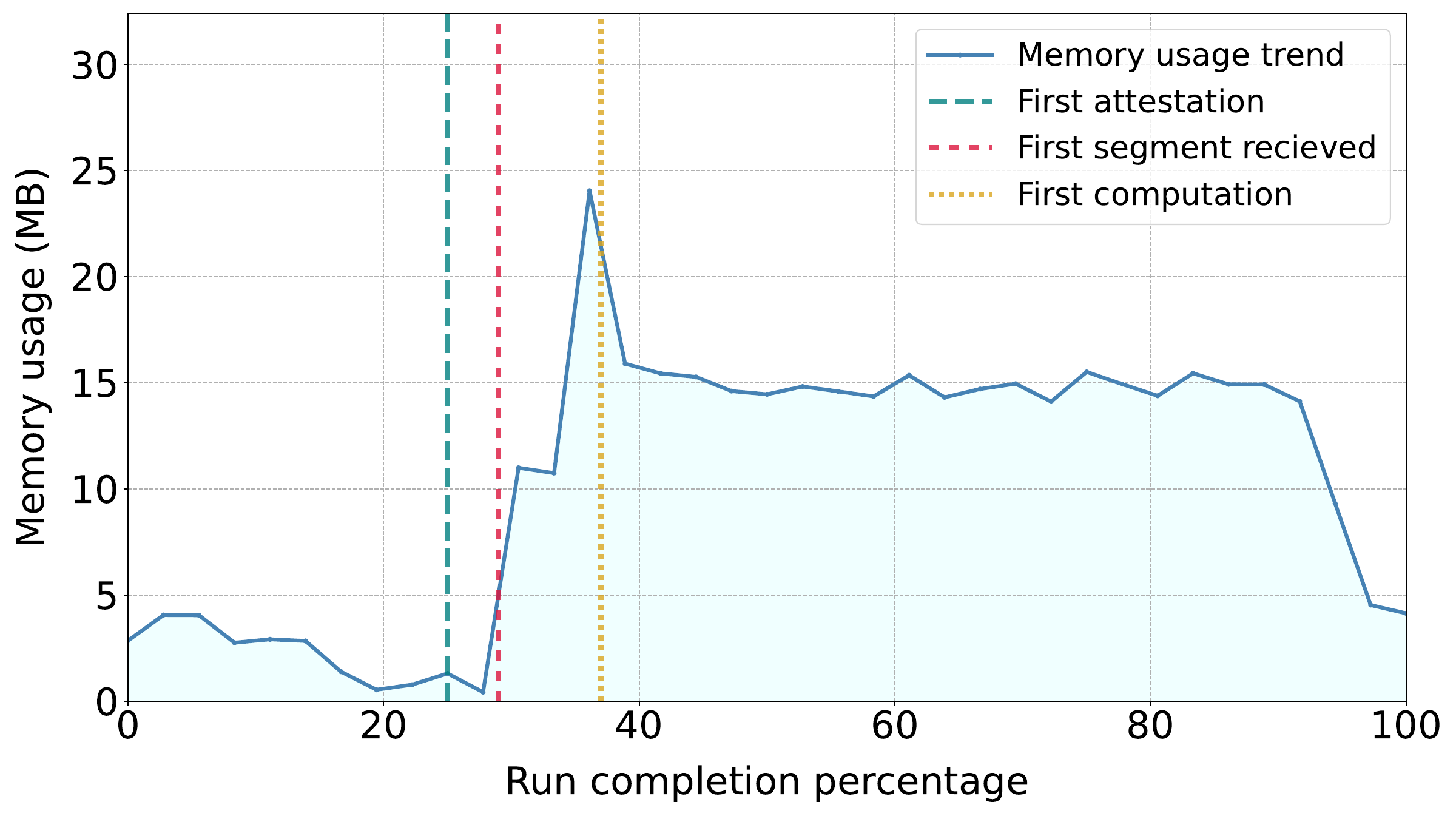}
  \caption{Memory usage with the computation phase}
  \label{snr_b}   
\end{subfigure}

\begin{subfigure}{0.49\textwidth}   
  \centering      
  \includegraphics[width=\textwidth]{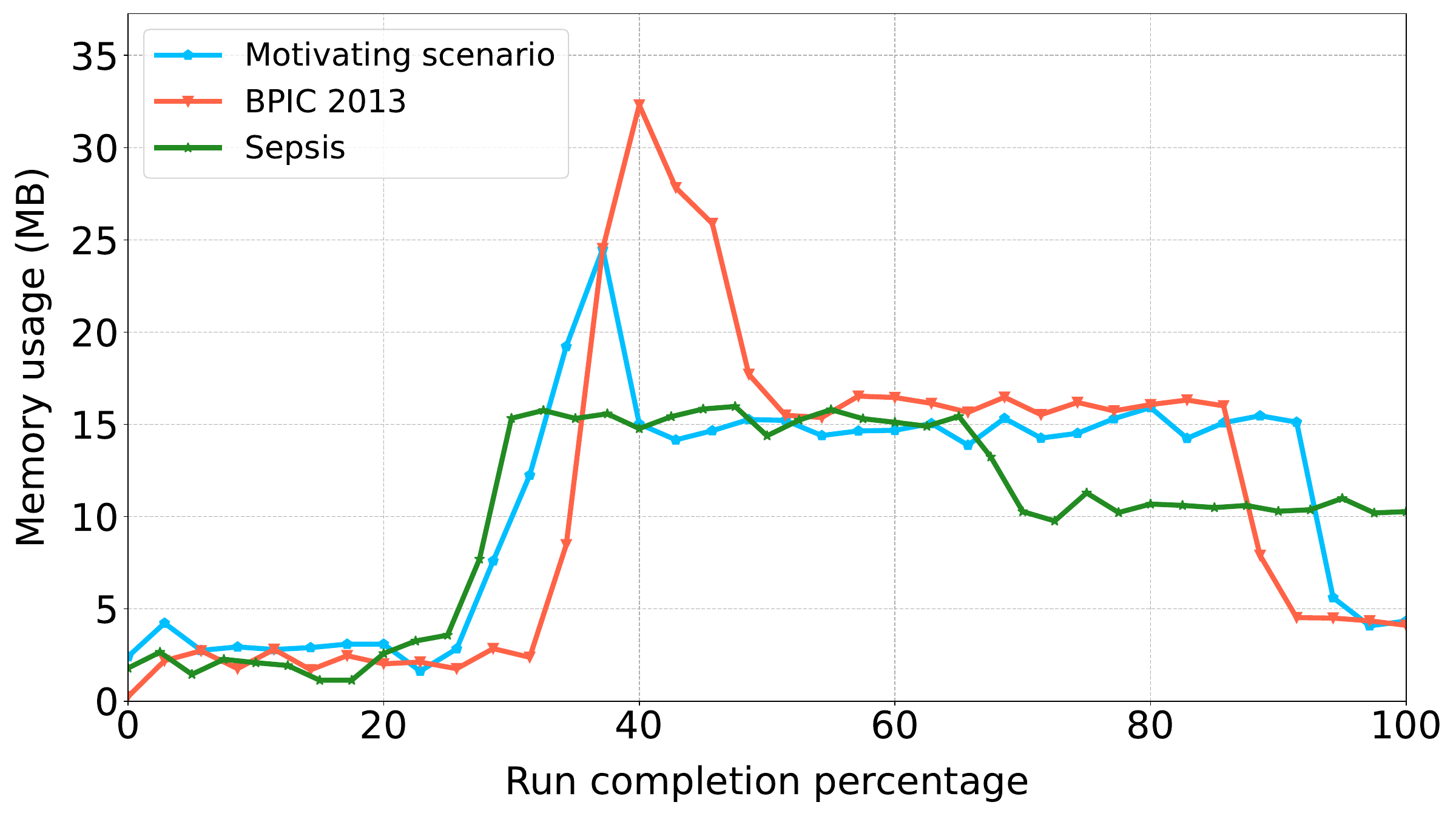}
  \caption{Memory usage with three event logs}
  \label{snr_c}
\end{subfigure}
\begin{subfigure}{0.49\textwidth}   
  \centering
  \includegraphics[width=\textwidth]{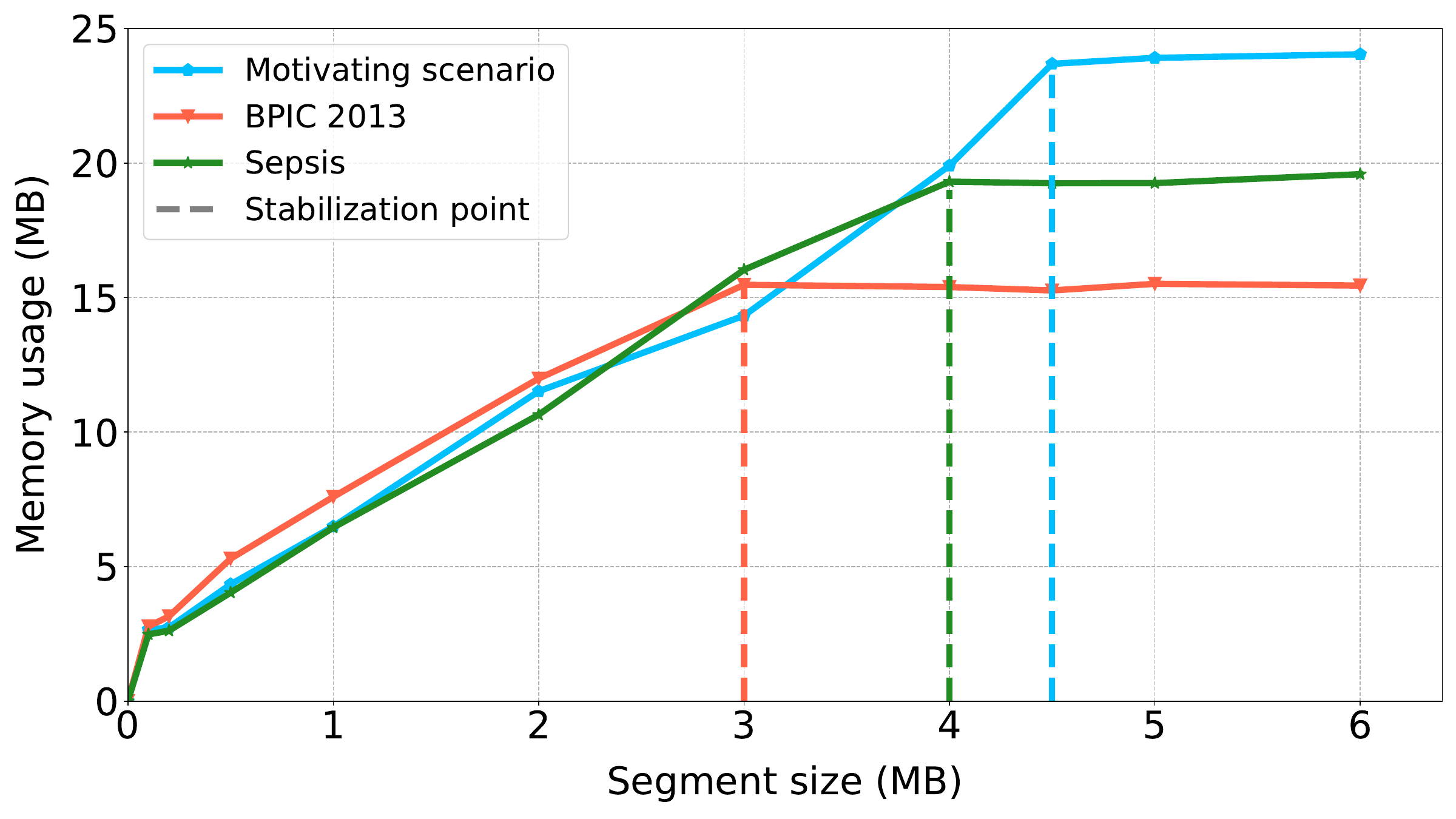}
  \caption{Segment size impact on memory usage}
  \label{snr_d}
\end{subfigure}
\caption{Memory usage test results}
\label{fig:memtest}
\end{figure}

\begin{figure}[t]
\begin{subfigure}{0.49\textwidth}   
  \centering
  \includegraphics[width=\textwidth]{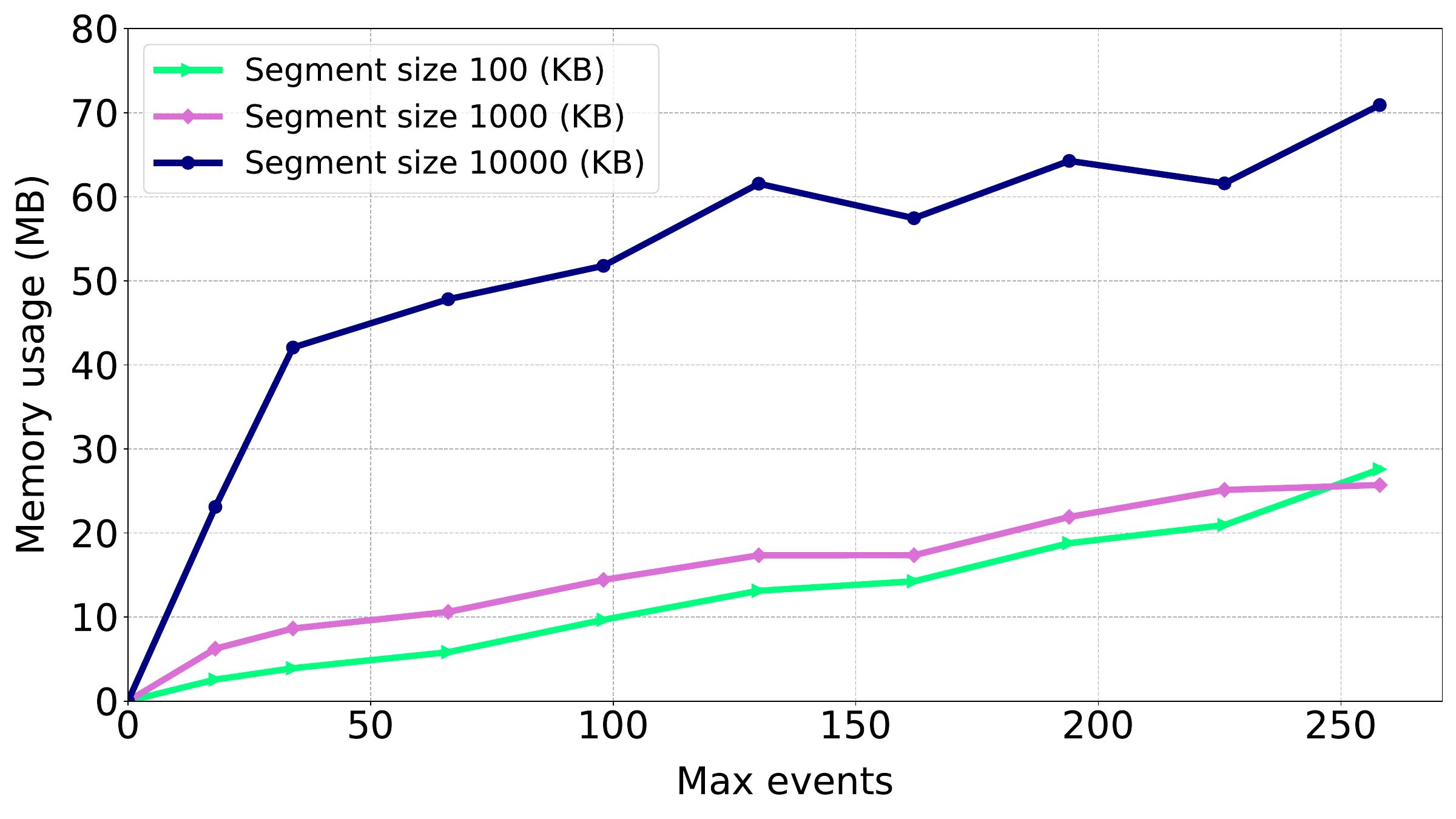}
  \caption{Results of scalability test \ref{test:events}}
  \label{fig:event_results}
\end{subfigure}
\begin{subfigure}{0.49\textwidth}   
  \centering
  \includegraphics[width=\textwidth]{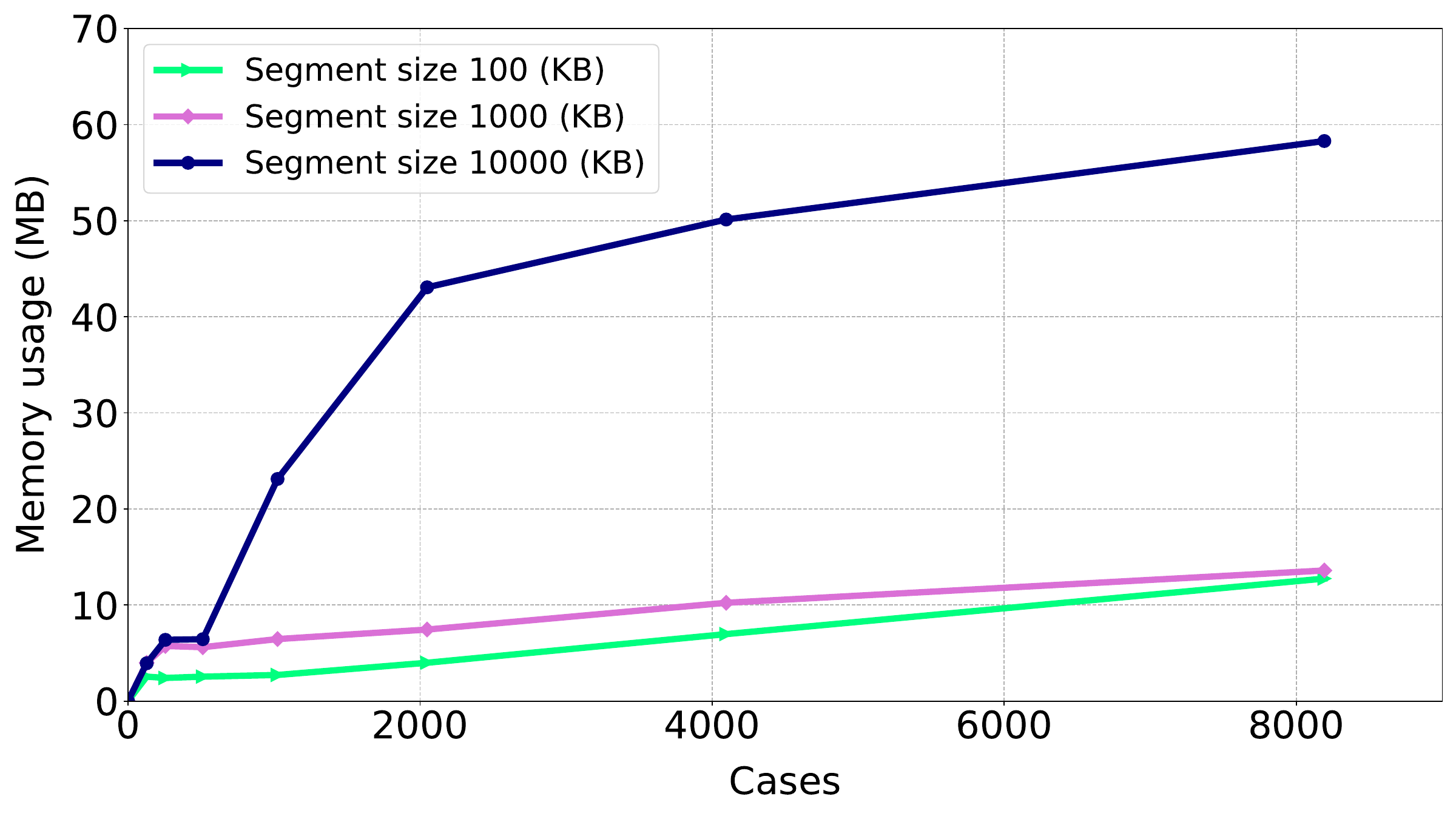}
  \caption{Results of scalability test \ref{test:cases}}
  \label{fig:cases_results}
\end{subfigure}

\begin{subfigure}{0.49\textwidth}   
  \centering
  \includegraphics[width=\textwidth]{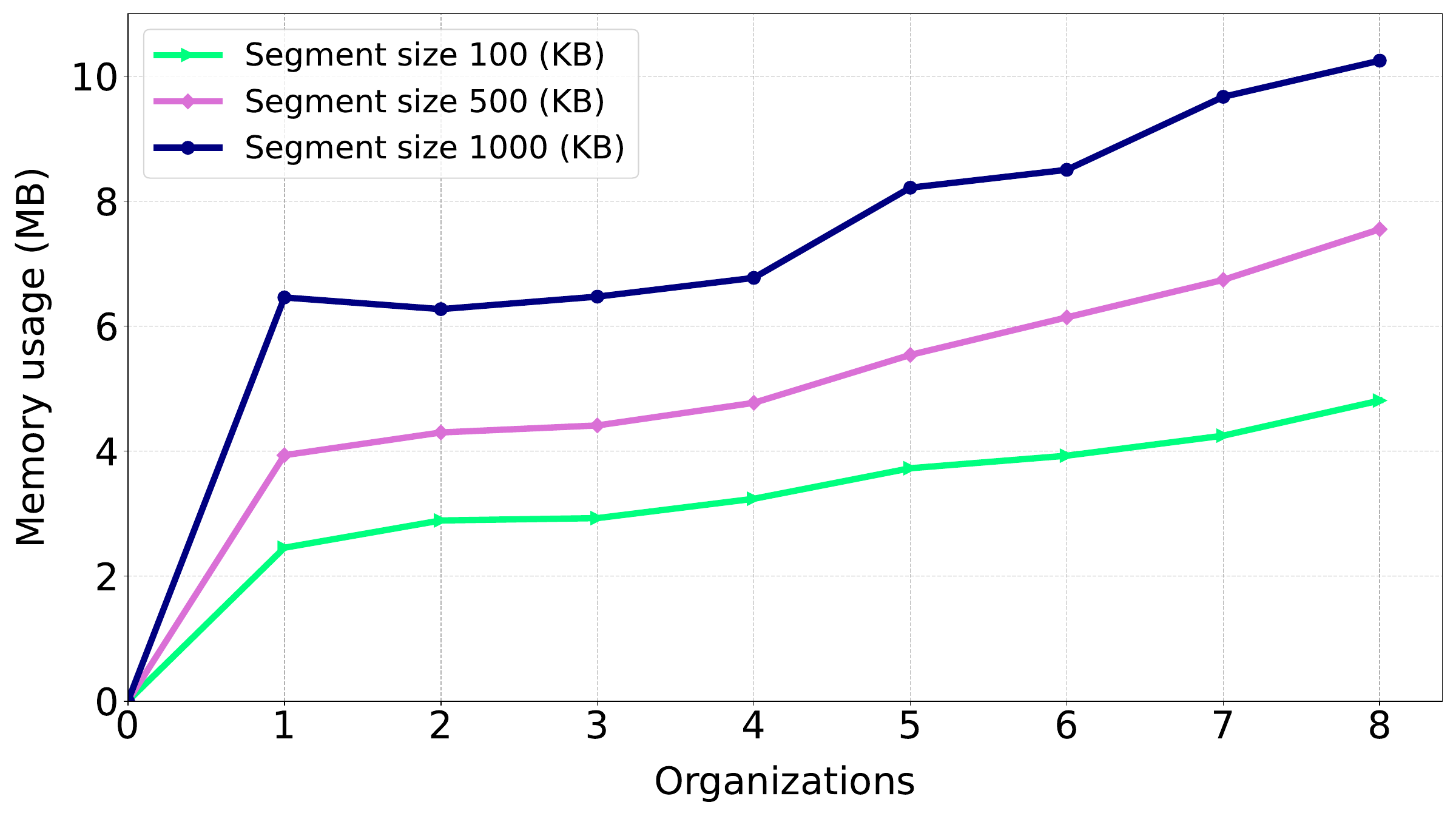}
  \caption{Results of scalability test \ref{test:organizations}}
  \label{fig:org_results}
\end{subfigure}
\begin{subtable}{0.49\textwidth}
  \centering
   \resizebox{0.97\textwidth}{!}{%
	\begin{tabular}{ll>{\hspace{1em}}cc>{\hspace{1em}}c}
	\toprule
	\textbf{Test} & \textbf{Seg.size} & 
	\textbf{\Rlin} & 
		\textbf{$\Slope$}& 
		$\Rlog$\\
		\midrule
		\multirow{3}{*}{\makecell[l]{Max\\events}} &
		100&0.9847& 0.0980&0.8291\\
		& 1000&0.9544&0.0821&0.9043\\
		& 10000&0.7357&0.1518&0.9386\\
		\midrule
		\multirow{3}{*}{\makecell[l]{Number\\of\\cases}} &
		100&0.9896&0.0013&0.6822\\
		& 1000&0.9629&0.0010&0.8682\\
		& 10000&0.7729&0.0068&0.9303\\
		\midrule
		\multirow{3}{*}{\makecell[l]{Provisioning\\organizations}} &
		100 &0.9770&0.3184&0.8577\\
		& 500 &0.9602&0.5174&0.7902\\
		& 1000 &0.9066&0.6102&0.6977\\
		\bottomrule
	\end{tabular}

    }
  \label{table:TestCoefficentTable}
  \caption{Scalability measurements}
\end{subtable}
\caption{Scalability test results}
\label{fig:scalabtest}
\end{figure}

\noindent\textbf{Memory usage.} \label{sec:evaluation:subsec:MemoryUsage} 
\Cref{snr_a,snr_b} display plots corresponding to the runtime space utilization of CONFINE (in MegaBytes). Differently from \cref{snr_b}, \cref{snr_a} excludes the computation stage by leaving the HeuristicsMiner inactive so as to isolate the execution from the mining-specific operations. 
The dashed lines mark 
the starting points for the remote attestation, data transmission and computation stages. 
We held the segment size ({\SegSize}) constant at \num{2} MegaBytes. 
We observe that the data transmission stage reaches the highest peak of memory utilization, 
which is then partially freed by the subsequent computation stage, steadily occupying memory space at a lower level. 
To verify whether this phenomenon is due to the synthetic nature of our simulation-based event log, we gauge the runtime memory usage of two public real-world event logs, too: Sepsis~\cite{seps}
and BPIC 2013~\cite{bpic2013}.
The characteristics of the event logs are summarized in \cref{tab:testedlogs}.
Since those are \textit{intra-organizational} event logs, we 
split the contents to mimic an \textit{inter-organizational} context.
In particular, we separated the Sepsis log based on the distinction between normal-care and intensive-care paths, as if they were conducted by two distinct organizations. Similarly, we processed the BPIC~2013 log to sort it out into the three departments of the Volvo IT incident management system. 
\Cref{snr_c} depicts the results.
\NewJ{%
We observe that the BPIC 2013 log demands the most memory during the initial stages, 
whereas the Sepsis log 
is associated with the least expensive run,
but the polylines exhibit a matching shape with our synthetic dataset.
}%
To verify whether these trends are affected by the dimension of the exchanged data segments, we conducted an additional test to examine memory usage as the {\SegSize} varies. 
Notably, the polylines displayed in \cref{snr_d} indicate a linear increment of memory occupation until a breakpoint is reached. After that, the memory in use is steady. These points, marked by vertical dashed lines, indicate that 
the {\SegSize} value that allows the providers to send their whole log partition in a single segment.

\noindent\textbf{Scalability.}
To examine the scalability of the \Compo{Secure Miner}, we focus on its capacity to efficiently manage an increasing workload in the presence of limited memory resources
\NewJ{%
(as it is the case with TEEs).
}%
We set three distinct test configurations 
by varying our motivating scenario log. In particular, we considered
\begin{inparaenum}[(I)]
	\item the maximum number of events per case,
        \label{test:events}
	\item the number of cases $|{\CIdU}|$, and
        \label{test:cases}
	\item the number of provisioning organizations $|{\OrgU}|$
        \label{test:organizations}
\end{inparaenum}
as independent integer variables. To conduct the test on the maximum number of events, we added a loop back from the final to the initial activity of the process model, progressively increasing the number of iterations $2 \leqslant \,x_\circlearrowleft\, \leqslant 16$ at a step of \num{2}, resulting in $18+16\cdot(x_\circlearrowleft-1)$ events. Concerning the test on the number of cases, we simulated additional process instances so that $|{\CIdU}| = 2^{x_{\CId}}$ having $x_{\CId} \in \{7,8,\ldots,13\}$. Finally, for the assessment of the number of organizations, the test necessitated the distribution of the process model activities' into a variable number of pools, each representing a different organization ($|{\OrgU}| \in \{1,2,\ldots,8\}$).
We parameterized the above configurations with three segment sizes (in KiloBytes): $\SegSize \in \{100,1000,10000\}$ for tests \ref{test:events}~and~\ref{test:cases}, and $\SegSize \in \{100,500,1000\}$ for test~\ref{test:organizations} (the range is reduced without loss of generality to compensate the partitioning of activities into multiple organizations). To facilitate a more rigorous interpretation of the output trends across varying {\SegSize}s, we employ 
two well-known statistical measures. As a primary measure of goodness-of-fit, we employ the coefficient of determination {\RCoefficent}~\cite{barrett1974coefficient}, which assesses the degree to which the observed data adheres to the linear ({\Rlin}) and logarithmic ({\Rlog}) regressions derived from curve fitting approximations. To delve deeper into the analysis of trends exhibiting a high {\Rlin}, we consider the slope {\Slope} of the approximated linear regression~\cite{altman2015simplelinearregression}. 

\hyperref[table:TestCoefficentTable]{Table 9(d)} lists the obtained measurements, which we use to elucidate the observed patterns. \Cref{fig:event_results} depicts the results of test \ref{test:events}, focusing on the increase of memory utilization when the number of events in the logs grows. We observe that the memory usage trends for {\SegSize} set to \num{100} and \num{1000} (depicted by green and lilac lines, respectively) are almost superimposable, whereas the setting with $\SegSize = 10000$ (blue line) exhibits significantly higher memory usage. With {\SegSize} assigned with \num{100} and \num{1000}, {\Rlin} approaches \num{1}, 
signifying an almost perfect approximation of the linear relation. 
With these settings, {\Slope} is very low 
yet higher than \num{0}, thus indicating that memory usage is likely to continue increasing as the number of maximum events grows. The configuration with $\SegSize = 10000$ yields a higher {\Rlog} value, 
thus suggesting a logarithmic trend, hence 
a greater likelihood of stabilizing memory usage growth rate as the number of maximum events increases. 
In  \cref{fig:cases_results}, we present the results of test \ref{test:cases}, assessing the impact of the number of cases on memory consumption. As expected, the configurations with {\SegSize} set to \num{100} and \num{1000} demand lower memory than settings with $\SegSize=10000$. The {\Rlin} score when {\SegSize} is assigned with \num{100} and \num{1000} 
indicate a strong linear relationship between the dependent and independent variables compared to the trend with $\SegSize = 10000$, which is better described by a logarithmic regression (${\Rlog} = \num{0.9303}$). 
Differently from test \ref{test:events}, the {\Slope} score associated with the linear approximations with {\SegSize} set to \num{100} and \num{1000} 
approaches \num{0}, indicating that the growth rate of memory usage as the number of cases increases is negligible.
In \cref{fig:org_results}, we present the results of test~\ref{test:organizations}, on the relation between the number of organizations and memory usage. The chart shows that memory usage trends increase as provisioning organizations increase for all three segment sizes. The {\Rlin} values for the three {\SegSize}s are very high, 
indicating a strong positive linear correlation. 
The test with $\SegSize = 100$ exhibits the slowest growth rate, as corroborated by the lowest {\Slope} (\num{0.3184}). 
For the configuration with $\SegSize = 500$, the memory usage increases slightly faster (${\Slope} = {0.5174}$). With $\SegSize = 1000$, the overall memory usage increases significantly faster than the previous configurations (${\Slope} = 0.6102$). We derive from these findings that the \Compo{Secure Miner} may encounter scalability issues when handling settings with a large number of provisioning organizations. Further investigation is warranted to determine the precise cause of this behavior and identify potential mitigation strategies. 

In the next section, we discuss other future endeavors stemming from our work.

\section{Conclusion and Future Work}
\label{sec:conclusion}
\NewJ{
In this paper, we described CONFINE, a
decentralized 
approach 
to process mining. 
Based upon TEEs, it guarantees the secrecy and confidentiality of data transmitted to and elaborated by processing nodes outside the perimeter of the event log providers.
Our research can spur a number of future investigations and improvements in the field. 
First, we aim to enhance our solution by readjusting it to the relaxation of underlying assumptions we made, including 
fair conduct by data provisioners, 
the absence of injected or maliciously manipulated event logs, the 
exchange of messages through reliable communication channels where no loss or bit corruption occurs,
and the 
existence of a universal clock for timestamps. 
}%
\NewJ{%
Also, we are extending our analysis with formal proofs of soundness and completeness of the protocol.
}
Our future work encompasses the integration of usage control policies that specify rules on event logs' utilization, too. We plan to design enforcement and monitoring mechanisms to achieve this goal following the principles adopted in \cite{basile2023blockchain,basile2023solid}. 
\NewJ{%
We remark that a possibly severe threat to data secrecy lies in the reconstruction of the original input information back from the mining output. Keeping this aspect in mind is crucial to determine the mining algorithm to be embedded in the \Compo{Secure Miner}. Studies in this regard have been conducted, among others, in~\cite{DBLP:conf/caise/VoigtFJKTMLW20,DBLP:journals/is/FahrenkrogPetersenKAW23}. Integrating the proposed recommendations with CONFINE paves the path for future investigations.
Finally, we acknowledge that the focus of our implementation is on a specific process discovery task. Nevertheless, our approach has the potential to seamlessly cover a wider array of discovery techniques as well as other process mining functionalities like conformance checking and performance analysis. Showing their integrability with our approach, and drawing guidelines on the use of different algorithms, are research directions we plan to follow.
}

\noindent\textbf{Acknowledgments.} The authors thank Giuseppe Ateniese for the fruitful discussion and insights.
This work was partly funded by MUR under PRIN grant B87G22000450001 (PINPOINT), the Latium Region under PO~FSE+ grant B83C22004050009 (PPMPP), and Sapienza University of Rome under grant RG123188B3F7414A (ASGARD). 

\vspace{-2ex}
\bibliographystyle{splncs04}
\bibliography{bibliography}

\end{document}